\documentclass[preprint,showpacs,preprintnumbers,amsmath,amssymb]{revtex4}
\usepackage{graphicx}
\begin{document}
\title{An analytic method of describing  $R$-related quantities in QCD}
\author{K.A. Milton}
\altaffiliation{On leave from the Oklahoma Center for High Energy Physics
and the Homer L. Dodge Department of Physics and
Astronomy, University of Oklahoma, Norman, OK 73019 USA}
\email{milton@nhn.ou.edu}
\affiliation{Department of Physics, Washington University, St. Louis,
MO 63130 USA}

\author{I.L. Solovtsov}
\altaffiliation[Also at ]{Bogoliubov Laboratory of Theoretical Physics,
Joint Institute for Nuclear Research, Dubna, 141980 Russia.}
\email{solovtsov@gstu.gomel.by, solovtso@thsun1.jinr.ru}
\author{O.P. Solovtsova}
\affiliation{International Center for Advanced Studies, Gomel State Technical University, Gomel, 246746 Belarus}

\date{\today}

\begin{abstract}
A model based on the analytic approach to QCD, involving a summation of
threshold singularities and taking into account
the nonperturbative character of the light quark masses, is applied to find
hadronic contributions to different
physical quantities. It is shown that the suggested model allows us to
describe well such objects as the hadronic
contribution to the anomalous magnetic moment of the muon, the ratio of
hadronic to leptonic $\tau$-decay widths in
the vector channel, the Adler $D$-function, the smeared $R_\Delta$-function,
and the hadronic contribution to the evolution of the fine structure constant.
\end{abstract}

\maketitle
\section{Introduction}
A comparison of QCD theoretical results with experimental data is often
based on the concept of
quark-\-had\-ron duality, which establishes a bridge between quarks and gluons,
a language of theoreticians, and
real measurements with hadrons performed by experimentalists. The idea of
quark-\-had\-ron duality was formulated in
the paper by Poggio, Quinn, and Weinberg \cite{PoggioQW76} as follows:
Inclusive hadronic cross sections, once they are
appropriately averaged over an energy interval, must approximately coincide
with the corresponding quantities
derived from the quark-gluon picture. For many physical quantities and
functions the corresponding interval of
integration involves an infrared region and in this case nonperturbative
effects may play an important role in their description.

In this paper we consider the following quantities and functions.
\begin{itemize}
\item
The ratio of hadronic to leptonic $\tau$-decay widths in the vector channel:
\begin{equation}\label{R_tau-V}
R_{\tau}^{V}=R^{(0)}\,\int\limits_0^{M_{\tau}^2}\frac{ds}{M_{\tau}^2}
\left(1-\frac{s}{M_{\tau}^2} \right)^2
\left(1+\frac{2s}{M_{\tau}^2}\right){R}(s);
\end{equation}
\item
the so-called ``light'' Adler function, which is constructed from
$\tau$-decay data \cite{MSS-Adler-funct:01}:
\begin{equation}\label{D-funct-def}
D(Q^2)=-Q^2\,\frac{d\,\Pi(-Q^2)}{d\,Q^2}=Q^2\,\int\limits_0^\infty\,ds\,
\frac{R(s)}{(s+Q^2)^2};
\end{equation}
\item
the smeared $R_\Delta$ function \cite{PoggioQW76}:
\begin{equation} \label{R(s)_Delta_2}
R_{\Delta}(s)=\frac{\Delta}{\pi} \int\limits_0^\infty ds'
\frac{R(s')}{(s-s')^2+\Delta^2};
\end{equation}
\item
the hadronic contribution to the anomalous magnetic moment of the muon
(in the leading order in electromagnetic coupling constant):
\begin{equation}\label{a_mu-QCD-def}
a_\mu^{\rm{had}}=\frac{1}{3}\left(\frac{\alpha}{\pi}\right)^2\,
\int\limits_{0}^\infty\frac{ds}{s}\,K(s)\,R(s),
\end{equation}
where $K(s)$ is the vacuum polarization factor given by (\ref{K(s)-integral})
below;
\item and
the strong interaction contribution to the running of the
fine structure constant:
\begin{equation}\label{Delta-alpha-R}
\Delta\alpha_{\rm{had}}^{(5)}(M_Z^2)=-\frac{\alpha(0)}{3\pi}\,M_Z^2\,
{\cal{P}}\,\int\limits_0^\infty
\frac{ds}{s}\,\frac{R(s)}{s-M_Z^2}\,.
\end{equation}
\end{itemize}

A common feature of all these quantities and functions is that they are defined
through the function $R(s)$, the
normalized hadronic cross-section, integrated with some other
functions. By definition, all these quantities
and functions include an infrared region as a part of the interval of
integration and, therefore, they cannot be directly
calculated within perturbative quantum chromodynamics (pQCD).

The method that we use here to describe the quantities and functions
mentioned above is based on the analytic approach to
QCD suggested in~\cite{SS:96,SS:97}. The analytic approach allows one to
describe self-consistently the timelike
region \cite{MS-time,Milton-Solovtsova:98}, which is represented in
the integration
in Eqs.~(\ref{R_tau-V})--(\ref{Delta-alpha-R}).
It incorporates
the required analytic properties and leads to an integral representation for
$R(s)$.  We formulate a model that also incorporates a summation of
threshold singularities \cite{Milton-Solovtsov_ModPL:01} and takes into account
the nonperturbative character of the light quark masses.

\section{Method and basic relations}

\subsection{Analytic perturbation theory}
Analytic perturbation theory (APT) \cite{MSS:97} is based on the analytic
approach to QCD. In this approach, in
contrast to the behavior of the perturbative running coupling, the Euclidean
analytic coupling has no unphysical
singularities. The ghost pole and corresponding branch points, which appear
in higher orders, are absent. APT
preserves the correct analytic properties of such important objects as
the two-point correlation function and also
provides a well-defined algorithm for calculating higher-loop corrections
\cite{Solovtsov-Shirkov:99-TMP}. In
APT, processes with typical spacelike and timelike momenta are described
self-consistently
\cite{MS-time,Milton-Solovtsova:98,DV:01-Eur} and, for example,
inclusive $\tau$-decay can be described equivalently either in terms of
Minkowskian or Euclidean variables \cite{MSS:97}. In the framework of APT,
the theoretical ambiguity
associated with the  choice of renormalization scheme is dramatically reduced.

In the APT scheme the QCD contributions $d(z)$ and $r(s)$ to
the functions $D\propto1+d$ and $R\propto1+r$,
respectively, are expressed in terms of the effective spectral function
$\rho(\sigma)$ as \cite{SS:96,SS:97,MS-time}
\begin{equation} \label{drho-rrho}
d(z)=\frac{1}{\pi}\int^{\infty}_0 \frac{d\sigma}{\sigma - z}\, \rho(\sigma)\,,
\qquad
r(s)=\frac{1}{\pi}\int^{\infty}_{s}\frac{d\sigma}{\sigma} \rho(\sigma)\, .
\end{equation}
The APT spectral function $\rho(\sigma)$ is defined as the imaginary part of
the perturbative function $d_{\rm pt}(z)$
and in the third order can be written in the form
$\rho(\sigma)=\varrho_0 (\sigma)+d_1 \varrho_1
(\sigma)+d_2\varrho_2 (\sigma),$ where
$\varrho_k (\sigma)={\rm Im}[a_{\rm pt}^{k+1}(\sigma+i\epsilon)]$,
$a=\alpha_s/\pi$, and $d_k$ are the coefficients of the perturbative
expansion of the $D$-function.

The function $\varrho_0 (\sigma)$ defines the analytic spacelike,
${\cal{A}}(z)$, and timelike,
${\mathfrak{A}}(s)$, running couplings as follows
\begin{equation}\label{a_t-a_s}
{\cal{A}}(z)=\frac{1}{\pi}\int_0^\infty\frac{d\sigma}{\sigma-z}\,
\varrho_0(\sigma)\,, \qquad
{\mathfrak{A}}(s)=\frac{1}{\pi}\int^{\infty}_{s}\frac{d\sigma}{\sigma}
\varrho_0(\sigma).
\end{equation}
As has been argued from general principles in~\cite{MS99}, the behavior of
these couplings cannot be the same, i.e., they
cannot be symmetrical in the spacelike and timelike domains,
${\cal{A}}(-z)\ne{\mathfrak{A}}(z)$. The analytic and perturbative
couplings have been compared in~\cite{Milton-Solovtsova:98}.
In analyzing hadronic processes with characteristic
spacelike and timelike momenta, it is necessary to take into account
this lack of symmetry between the behavior
of the running coupling in the Euclidean and Minkowskian regions.

The analytic running coupling has no unphysical
singularities and possesses the correct analytic properties, arising from
K\"all\'en-Lehmann analyticity reflecting
the general principles of the theory. The one-loop APT result
is~\cite{SS:96,SS:97,MS-time}
\begin{equation} \label{a1}
{\cal{A}}^{(1)}(z)=a^{(1)}_{\rm pt}(z)+\frac{4}{\beta_0}\frac{\Lambda^2}
{\Lambda^2+z},\qquad
{\mathfrak{A}}^{(1)}(s)=\frac{4}{\beta_0}\left[\frac{1}{2}
-\frac{1}{\pi}\arctan\frac{\ln( s/\Lambda^2)}{\pi} \right],
\end{equation}
where $a^{(1)}_{\rm pt}(z)=\bar{\alpha}_s(z)/\pi=4/\left[\beta_0
\ln(-z/\Lambda^2)\right]\,$ and $\beta_0=11-2f/3$
is the first coefficient of the renormalization group $\beta$-function.

Both the couplings (\ref{a_t-a_s}) have the same infrared fixed point
${\cal{A}}(0)={\mathfrak{A}}(0)=4/\beta_0$.
This value is defined by the leading contribution (\ref{a1}) and
is not altered by higher-order corrections. The
regular behavior in the infrared region of ${\cal{A}}^{(1)}(z)$ is provided
by the power term in (\ref{a1}) which
is invisible in the perturbative expansion. The reason for the
regularity of the coupling
${\mathfrak{A}}^{(1)}(s)$ is connected with the summation of the so-called
$\pi^2$-terms that play an important role in
analyzing various hadronic processes
\cite{Radyushkin82,KP82,Bjorken89,GKL91,DV:01-Eur,kataev95}.

\subsection{Resummation of threshold singularities}
In describing a charged particle-antiparticle system near threshold, it is
well known from QED that the so-called
Coulomb resummation factor plays an important role. This resummation,
performed on the basis of the nonrelativistic
Schr\"odinger equation with the Coulomb potential $V(r)=-\alpha/r$,
leads to the Som\-mer\-feld-\-Sakha\-rov
$S$-factor~\cite{Sommerfeld,Sakharov}. In the threshold region one cannot
truncate the perturbative series and the
$S$-factor should be taken into account in its entirety. The $S$-factor
appears in the parametrization of the
imaginary part of the quark current correlator, which can be approximated
by the Bethe-Salpeter amplitude of the two
charged particles, $\chi_{\rm{BS}}(x=0)$~\cite{BarbieriCR73}.
The nonrelativistic replacement of this amplitude by
the wave function, which obeys the Schr\"odinger equation with the Coulomb
potential, leads to the appearance of the
resummation factor in the parametrization of the $R(s)$-function discussed
above.

For a systematic relativistic analysis of quark-antiquark systems,
it is essential from the very beginning to have a
relativistic generalization of the $S$-factor. A new form for this
relativistic factor in the case of QCD has been
proposed in~\cite{Milton-Solovtsov_ModPL:01}
\begin{equation}\label{S-factor-relativistic}
S(\chi)=\frac{X(\chi)}{1-\exp\left[-X(\chi)\right]}\, ,
\quad\quad X(\chi)=\frac{\pi\,\alpha}{\sinh\chi}\, ,
\end{equation}
where $\chi$ is the rapidity which related to $s$ by
$2m\cosh\chi=\sqrt{s}$, $\alpha\to 4\alpha_s/3$ in QCD. The
function $X(\chi)$ can be expressed in terms of $v=\sqrt{1-4m^2/s}$:
$X(\chi)=\pi\alpha\sqrt{1-v^2}/v$. The
relativistic resummation factor (\ref{S-factor-relativistic}) reproduces
both the expected nonrelativistic and
ultrarelativistic limits and corresponds to a QCD-like Coulomb potential.
Here we consider the vector channel for
which a threshold resummation $S$-factor for the s-wave states is used. For
the axial-vector channel the $P$-factor
is required. The corresponding relativistic factor has recently been found
in~\cite{SS-Chernichenko:04}.

To incorporate the quark mass effects one usually uses the approximate
expression proposed in
\cite{App-Politzer75,PoggioQW76} above the quark-antiquark threshold
\begin{equation}\label{R-appr1}
{\cal{R}}(s)=T(v)\,\left[1+g(v)r(s)\right]\,,
\end{equation}
where
\begin{eqnarray}\label{vTg}
T(v)=v\frac{3-v^2}{2}\, ,\quad
g(v)=\frac{4\pi}{3}\left[\frac{\pi}{2v}-\frac{3+v}{4}
\left(\frac{\pi}{2}-\frac{3}{4\pi} \right) \right]\, , \quad
v_f=\sqrt{1-\frac{4m_f^2}{s}}\, .
\end{eqnarray}
The function $g(v)$ is taken in the Schwinger approximation \cite{Schwinger70}.

One cannot directly use the perturbative expression for $r(s)$ in
Eq.~(\ref{R-appr1}), which contains unphysical
singularities, to calculate, for example, the Adler $D$-function.
Instead, one can use the APT representation for $r(s)$. The
explicit three-loop form for $r_{\rm APT}(s)$ can be found in
\cite{MSS-Adler-funct:01}. Besides this replacement,
one has to modify the expression (\ref{R-appr1}) in such a way as to take into
account summation of an arbitrary number
of threshold singularities. Including the threshold resummation factor
(\ref{S-factor-relativistic}) leads to
the following modification of the expression (\ref{R-appr1})
\cite{MSS-Adler-funct:01,SSS_JETPh:01} for a particular quark flavour $f$
\begin{align}\label{R-R_0R_1}
{\cal{R}}_f(s)&=\left[R_{0,f}(s)+R_{1,f}(s)\right]\Theta (s-4m_f^2), \\
R_0(s)&=T(v)\,S(\chi), \qquad R_1(s)=T(v)\left[\,r_{\rm{APT}}(s)\,g(v)-\frac{1}{2}X(\chi)\,\right]. \nonumber
\end{align}
The usage of the resummation factor (\ref{S-factor-relativistic})
reflects the assumption that the coupling is taken in the $V$
renormalization scheme. To avoid double counting, the function $R_1$
contains the subtraction of $X(\chi)$. The
potential term corresponding to the $R_0$ function gives the principal
contribution to ${\cal{R}}(s)$, the correction
$R_1$ amounting to less than twenty percent for the whole energy interval
\cite{SS-Chernichenko:04}.

\subsection{Quark masses}
\label{sec:qm}
The following considerations suggest the behavior of the mass function of the
light quarks in the infrared
region. A solution of the Schwinger-Dyson equations
\cite{Roberts-Schmidt:00,Fisher-Alkofer:02-03,
Aguilar-Nesterenko-Papavassiliou:05} demonstrates a fixed infrared
behavior of the invariant charge and the quark mass function. The mass function
of the light quarks at small momentum looks
like a plateau with a height approximately equal to the constituent mass,
then with increasing momentum the mass function rapidly decreases
and approaches the small current mass.

This behavior can be understood by using the concept of the dynamical quark
mass. This mass has an essentially
nonperturbative nature. Its connection with the quark condensate has been
established in~\cite{Politzer-masses:76}.
By using an analysis based on the Schwinger-Dyson equations a similar
relation has been found
in~\cite{KP-Dynamical:82}. It has been demonstrated in
\cite{Elias-Scadron:84} that on the mass-shall one has a
gauge-independent result for the dynamical mass
\begin{equation} \label{m3-qq}
m^3=-\frac{4}{3}\pi\alpha_s \langle0\vert\bar{q}\,q\vert0\rangle.
\end{equation}
A result obtained in \cite{Reinders-Stam:87} demonstrates the
step-like behaviour of the mass function. The height $m$ of  the
plateau is given by the quark condensate (\ref{m3-qq}).
\begin{figure}[htb]
\centering\includegraphics[width=0.5\textwidth]{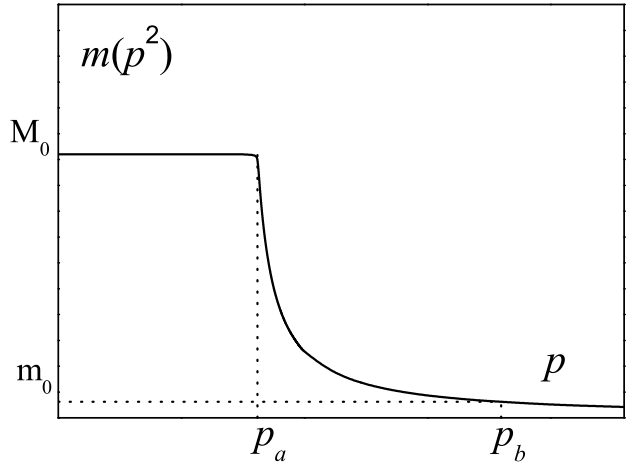} 
\vspace*{-0.4cm}\caption{Effective quark mass.} \label{dyn-mass-1}
\end{figure}
According to these results it is reasonable to assume that at small $p^2$
the function $m(p^2)$ is rather smooth
(nearly constant). In the region $p^2>1$--$2$~GeV the principal behavior
of the function $m(p^2)$ is defined by
perturbation theory with the renormalization group improvement.

\begin{table}[htb]
  \centering\caption{Typical values of $m_0^f$ and $M_0^f$.}\label{m_0-values}
\begin{tabular}{|c|c|c|c|c|c|c|} \hline
$f$& $u$& $d$       & $s$     & $c$     & $b$ &  $t$\\\hline
$~~m_0^f$ (GeV)~~&~ 0.004  &~ 0.007   &~ 0.130 &~ 1.35  &~ 4.4 ~&~ 174  \\  \hline
$~~M_0^f$ (GeV)~~&~ 0.260  &~ 0.260   &~ 0.450 &~ 1.35  &~ 4.4 ~&~ 174.0 \\  \hline
\end{tabular}
\end{table}

The following analysis was performed by using the model mass function $m(p^2)$
that is shown in Fig.~\ref{dyn-mass-1}. We take
the curve that connects the points $p_{\,a}$ and $p_{\,b}$ to have the form $A^3/(p^2-B^2)$.
The parameters $m_0$ are taken from the
known values of the running (current) masses at $p_{\,b}=2$~GeV. The quantities
considered here are not too sensitive to the
parameters of the heavy quarks and we take for $c$, $b$ and $t$ quarks
$m^f(p^2)=m_0^f=M_0^f=\mbox{const}$. The values of $m^f_0$
at $2$~GeV \cite{PDG:04} and typical values of $M_0^f$ are shown in
Table~\ref{m_0-values}.

\section{Physical quantities and functions generated by $R(s)$}
In this section we apply the model we have formulated to describe
the physical quantities and
functions connected with $R(s)$, described in the Introduction.
\subsection{Inclusive decay of the $\tau$-lepton}
The ratio of hadronic to leptonic $\tau$-decay widths in the vector channel is
expressed by Eq.~(\ref{R_tau-V}),
where $R^{(0)}={3}\,|V_{ud}|^2\,S_{\rm EW}/{2}$, $|V_{ud}|=0.9752\pm0.0007$ is
the CKM matrix element,
$S_{\rm{EW}}=1.0194\pm0.0040$ is the electroweak factor, and
$M_{\tau}=1776.99^{+0.29}_{-0.26}$~MeV is the mass of
the $\tau$-lepton \cite{PDG:04}. The experimental data obtained by the ALEPH
and OPAL collaborations for this ratio is \cite{ALEPH98,ALEPH02,OPAL99}:
$R_{\tau,V}^{\rm{ALEPH}}=1.775\pm0.017$, $R_{\tau,V}^{\rm{OPAL}}
=1.764\pm0.016$.

In our analysis we use the nonstrange vector channel spectral function
obtained by the ALEPH collaboration
\cite{ALEPH98} and keep in all further calculations the value
$R_{\tau,V}^{\rm{ALEPH}}$ as the normalization point.
The range of estimates are obtained by varying the quark masses in the
interval $M_0^{u,d}=260\pm10$~MeV (this band is fixed rather
definitely by the $D$-function considered below) and $M_0^{c}=450\pm100$~MeV.
The results for $R_\tau^V$ are given below.
\subsection{$D_V$-function}
The experimental information obtained by the ALEPH and OPAL collaborations
allows us to construct the nonstrange vector channel ``experimental''
$D$-function. Within the analytic approach this
function has been analysed in \cite{MSS-Adler-funct:01}. Here we improve our
method of constructing the ``light''
$D$-function by taking into account the global duality relation.
We demonstrate that this Euclidean object is useful
from the point of view of defining the effective masses of the light quarks.

In order to construct the Euclidean $D$-function (\ref{D-funct-def})
we use for $R(s)$ the following expression
\begin{equation}\label{R(s)-for-D}
R(s)=R^{\rm{expt}}(s)\,\theta(s_0-s)+R^{\rm{theor}}(s)\,\theta(s-s_0)\,.
\end{equation}
The continuum threshold $s_0$ we find from the global duality relation
\cite{Peris-Perrottet-Rafael:98}
\begin{equation}\label{global-duality}
\int\limits_0^{s_0}\,ds\,R^{\rm{expt}}(s)=\int\limits_0^{s_0}\,ds\,
R^{\rm{theor}}(s).
\end{equation}
This gives $s_0\simeq1.5$~GeV$^2$. The value of $s_0$ agrees with
the results of papers
\cite{Dorokhov:04,Rafael:94,Narison:04}. A similar value of the continuum
parameter is used in the QCD sum
rules~\cite{SVZ79,ReindersRY85}. Note, for some parameters there are two
possible solutions of the duality condition
(\ref{global-duality}). We exclude the second solution, $s_0\simeq2.5$~GeV$^2$,
at this stage of the analysis due to the requirement of describing,
in a self-consistent manner, different experimental data.

\begin{figure}[htb]
         \begin{minipage}[b]{0.49\textwidth}
\centering\includegraphics[width=0.99\textwidth]{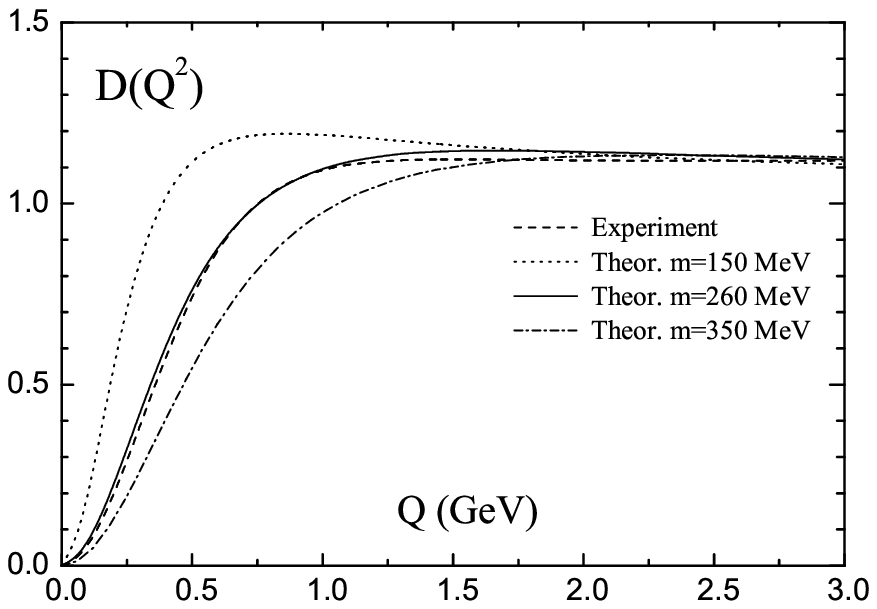}
         \end{minipage}%
\phantom{}\hspace{0.3cm}%
     \begin{minipage}[b]{0.489\textwidth}
\centering\includegraphics[width=0.956\textwidth]{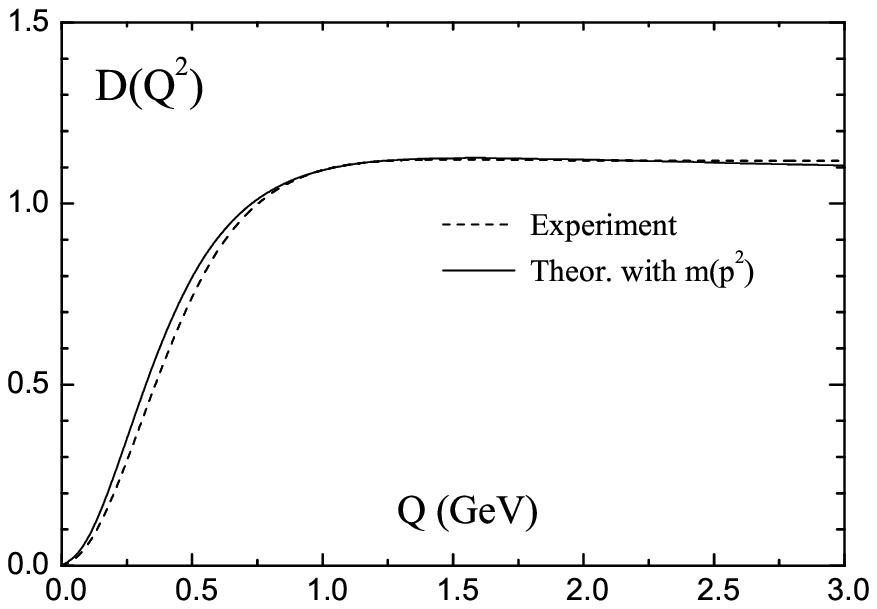} 
    \end{minipage}\\[-20pt]
             \begin{minipage}[t]{0.45\textwidth}
\vspace*{-0.1cm}\caption{$D$-function for $m=\mbox{const}$.} \label{d-funct}
        \end{minipage}%
\phantom{}\hspace{0.5cm}%
     \begin{minipage}[t]{0.45\textwidth}
\vspace*{-0.1cm}\caption{$D$-function  for $m=m(p^2)$.} \label{d-funct-var-masses}
    \end{minipage}      \end{figure}

The low energy $\tau$-data in the nonstrange vector channel results in
the curve for $D(Q^2)$ in Fig.~\ref{d-funct}.
In this figure we also plot three theoretical
curves corresponding to masses of the light quarks of
$150$, $260$ and $350$~MeV. Fig.~\ref{d-funct} demonstrates
that the shape of the infrared tail of the $D$-function is quite sensitive to
the value of the light quark masses.
Note the experimental $D$-function turns out to be
a smooth function without any trace of resonance structure. The
$D$-function obtained in Ref.~\cite{eidelman98} from the data for
electron-positron annihilation into hadrons also has a similar property.

\begin{figure}[bth]\begin{center}
\includegraphics[width=0.55\textwidth,height=0.55\textwidth]{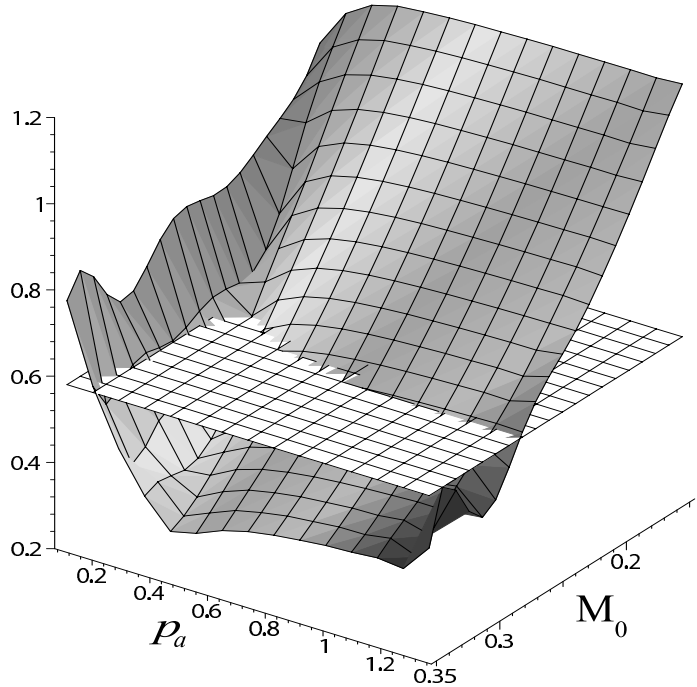}
\caption{$D$-function surface at $Q_0=0.5$ GeV vs. parameters $M_0$ and $p_{\,a}$.
The plane corresponds to $D_{\rm{expt}}(Q_0^2)$.}
\label{d-surface}
      \end{center} \end{figure}

The measured quantity $R_{\tau}^{V}$ defined in Eq.~(\ref{R_tau-V})
is less sensitive to $m_u$ and $m_d$ values than
the infrared tail of the $D_V$-function. Varying the light quark masses
over a wide range one finds $R_{\tau}^{V}=1.79$ for
$m_u=m_d=150$~MeV and $R_{\tau}^{V}=1.66$ for $350$~MeV.
The values of masses $m_u=m_d\simeq260$~MeV agree with
the experimental value $R_{\tau}^{V}=1.775\pm0.017$ \cite{ALEPH98}.
The values of the light quark masses are close to the constituent quark masses
and therefore incorporate nonperturbative
effects. These values are consistent with other results
\cite{Sanda:79,SakuraiSchTr:81,Solovtsov-Shirkov:99} and
with the analysis performed in
\cite{Dorokhov:04,Dorokhov:2005ff,Dorokhov:2003kf}.

A result for the $D$-function that is obtained by using the mass function
$m(p^2)$ with parameters defined in
Table~\ref{m_0-values} and $p_{\,a}=0.8$~GeV is shown in
Fig.~\ref{d-funct-var-masses}. Thus we obtain results that are
rather close to the results obtained for $m(p^2)=\mbox{const}=260$ MeV.
In Fig.~\ref{d-surface} we plot a 3-dimensional graph of
the function $D(Q_0^2)$ as function of the parameters $M_0$ and $a$.
The plane corresponds to the experimental value
$D_{\rm{expt}}(Q_0^2)=0.58$ at $Q_0=0.5$~GeV.
Fig.~\ref{d-surface} demonstrates that for $p_{\,a}>0.4$--$0.5$~GeV the
curve of intersection of the surface $D(Q_0^2;\,M_0,\,p_{\,a})$ and the plane is
approximately a straight line, corresponding to $M_0=260$ MeV. The
large $p_{\,a}$-limit reproduces the results with $m(p^2)=\mbox{const}$.

\subsection{Smeared $R_\Delta$-function}
To compare experimental and theoretical results from the point of view of
the quark-hadron duality, in
\cite{PoggioQW76} it was proposed to use the smeared function $R_\Delta(s)$.
Instead of the Drell ratio $R(s)$ defined in terms of the discontinuity
of the correlation function $\Pi(q^2)$ across the physical cut
\begin{equation}\label{R(s)}
R(s)=\frac{1}{2\pi\,i}\left[\Pi(s+i\epsilon)-\Pi(s-i\epsilon)\right],
\end{equation}
the smeared function $R_\Delta(s)$ is defined as
\begin{equation}\label{R(s)_Delta_1_b}
R_{\Delta}(s)=\frac{1}{2\pi\,{i}}\, \left[\,\Pi(s+{i}\Delta)-\Pi(s-{i}\Delta)\,
\right],
\end{equation}
with a finite value of $\Delta$ to keep away from the cut.
If $\Delta$ is sufficiently large and both the experimental
data and the theory prediction are smeared, it is possible to compare
theory with experiment.

\begin{figure}[htb]
         \begin{minipage}[b]{0.49\textwidth}
\centering\includegraphics[width=0.93\textwidth]{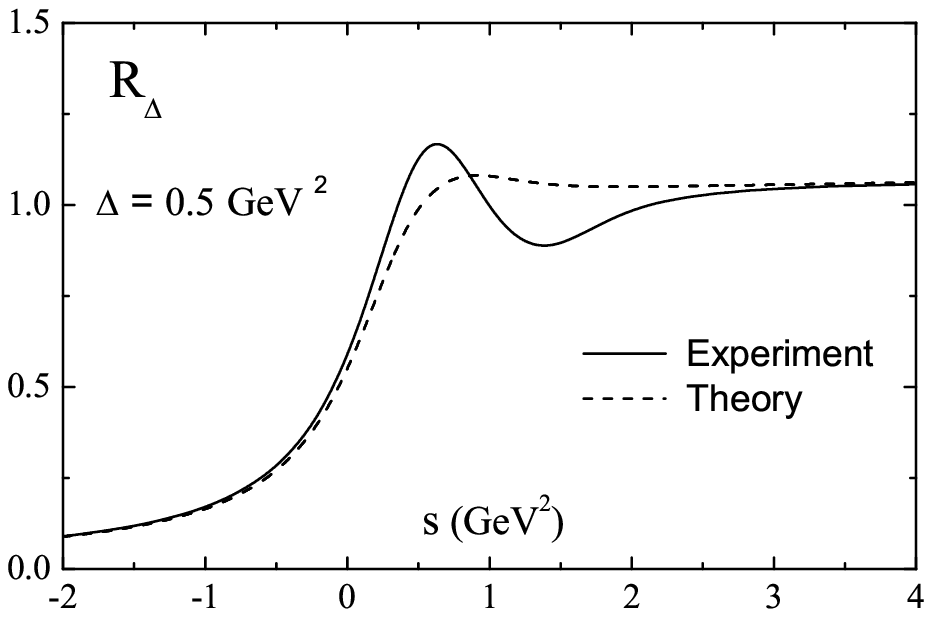}
         \end{minipage}%
     \begin{minipage}[b]{0.49\textwidth}
\centering\includegraphics[width=0.93\textwidth]{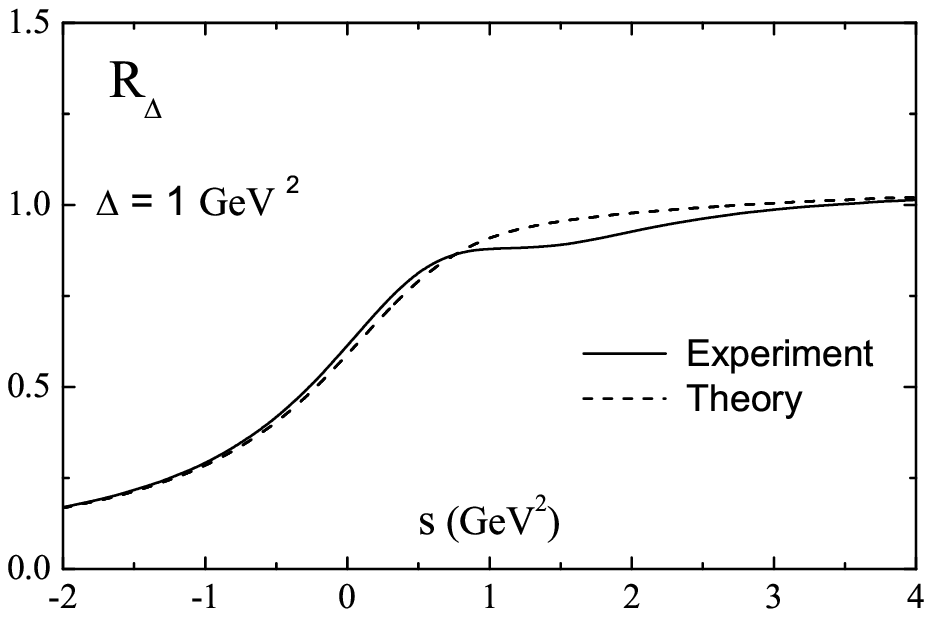} 
    \end{minipage}\\[-20pt]
             \begin{minipage}[t]{0.45\textwidth}
\vspace*{-0.0cm}\caption{Smeared function for $\Delta=0.5$ GeV$^2$.} \label{r-delta-05}
        \end{minipage}%
\phantom{}\hspace{0.5cm}%
     \begin{minipage}[t]{0.45\textwidth}
\vspace*{-0.0cm}\caption{Smeared function for $\Delta=1.0$ GeV$^2$.} \label{r-delta-10}
    \end{minipage}  \end{figure}

Equation~(\ref{R(s)_Delta_1_b}) and the dispersion relation for the
correlator $\Pi(q^2)$ give the representation
(\ref{R(s)_Delta_2}). Note that the smeared function $R_\Delta(s)$ is
defined both in the Minkowskian
region of positive $s$, where a trace of resonances still remains
for not too large $\Delta$, and in the Euclidean domain
of negative argument $s$, where like the Adler function
$D(Q^2)$ the function $R_\Delta(s)$ is smooth and monotone.

As with the Adler function we will construct the ``light'' experimental
function $R_\Delta(s)$. For this purpose we match
the experimental data taken with $s<s_0$ to the theoretical result taken
with $s>s_0$ as in (\ref{R(s)-for-D}). The
value $s_0\simeq1.6$~GeV$^2$ is found from the duality relation
(\ref{global-duality}).

For the charm region the value of $\Delta$ is about 3~GeV$^2$.
An adequate choice in the case of the light smeared
function is $\Delta\simeq0.5$--$1.0$ GeV$^2$. In Figs.~\ref{r-delta-05}
and \ref{r-delta-10} the experimental and
theoretical curves for $\Delta=0.5$ GeV$^2$, $\Delta=1.0$ GeV$^2$ and
$m=m(p^2)$ are shown. Let us emphasize that, for reasonable
values of $\Delta$, in the spacelike region ($s<0$) there is a good agreement
between data and theory starting from $s=0$.

\subsection{Hadronic contribution to $a_\mu$}
The hadronic contribution to the anomalous magnetic moment of the muon in the
leading order in the electromagnetic
coupling constant is defined by (\ref{a_mu-QCD-def}), where
$\alpha^{-1}=\alpha(0)^{-1}=137.035\,999\,11(46)$ \cite{PDG:04}, and
(see, for example, \cite{Schwinger70})
\begin{equation}\label{K(s)-integral}
K(s)=\int\limits_0^1\,dx\,\frac{x^2(1-x)}{x^2+(1-x)s/m_\mu^2}.
\end{equation}
The muon mass is $m_\mu=105.7$~MeV.

The expression (\ref{a_mu-QCD-def}) can be rewritten in terms of the
$D$-function
\begin{equation}\label{a_mu-QCD-via-D-funct}
a_\mu^{\rm{had}}=\frac{1}{3}\left(\frac{\alpha}{\pi}\right)^2\,\frac{1}{2}\,
\int\limits_{0}^1\frac{dx}{x}\,(1-x)(2-x)\,D\left(\frac{x^2}{1-x}\,m_\mu^2
\right).
\end{equation}

It is should be emphasized that the expressions (\ref{a_mu-QCD-def}) and
(\ref{a_mu-QCD-via-D-funct}) are equivalent
due to the analytic properties of the function $\Pi(q^2)$. If one uses a
method that does not maintain the required
properties of $\Pi(q^2)$, expressions (\ref{a_mu-QCD-def}) and
(\ref{a_mu-QCD-via-D-funct}) will no longer be equivalent
and will imply different results (see \cite{MS:02} for details).
This situation is similar to that which occurs in the analysis of inclusive
$\tau$-decay \cite{MSS:97}, where the initial integral,  performed over an
interval including a nonperturbative
region, for which a perturbative QCD calculation is not valid, is
transformed based on the analytic properties into a
contour representation. Within APT one is justified in doing this, and
can use equally well either the expression (\ref{a_mu-QCD-def}) or the
expression (\ref{a_mu-QCD-via-D-funct}).

\begin{table}[htb]
  \centering{\caption{Dependence of $a_\mu^{\rm{had}}$ on light quark
masses.}\label{mass-depend-tabl}
\begin{tabular}{|cc|cc|cc|}   \hline
\multicolumn{2}{|c|} {~~~} &\multicolumn{4}{|c|}{$a_\mu^{{\rm had}}
\times10^{10}$} \\  \cline{3-6}
\cline{1-2}
\multicolumn{2}{|c|} {$m_{q}$~(MeV)}
& \multicolumn{2}{|c|}{LO}   & \multicolumn{2}{|c|}{NNLO}  \\ \cline{3-6}
$~q=u,d$ &  $~q={s}$ & $~m_{q}=\mbox{const}$ & $m_{q}\neq\mbox{const}$
& $m_{q}=\mbox{const}$ & $m_{q}\neq\mbox{const}$ \\
\hline
250 & 400 & 736 &760 & 725 & 763 \\
250 & 500 & 716 &736 & 705 & 726 \\  \hline
260 & 400 & 691 &715 & 682 & 711 \\
260 & 500 & 671 &690 & 661 & 685  \\ \hline
\end{tabular}   } \end{table}

The value of $a_\mu^{\rm{had}}$ is not very sensitive to the values of the
heavy quark masses, which we take as given in
Table~\ref{m_0-values}. The relative contributions of $u$ and $d$ quarks
are about 72 and 19~\%, respectively. The
relative factor of 4 between $u$ and $d$ contributions is explained by the
ratio of quark charges. The relative
contribution of the $s$-quark to $a_\mu^{\rm{had}}$ is about 5--9~\% for
$M_0^s=400$--$500$~MeV. The contribution of the
$c$-quark is about $2\,\%$. Contributions of $b$ and $t$ quarks are very
small.

There is a significant dependence on the mass parameters of the light
quarks. This dependence we illustrate in
Table~\ref{mass-depend-tabl}. In our calculations we take into account the
matching conditions at quark thresholds
according to the procedure described in~\cite{Milton-Solovtsova:98}.
The mass parameters of $u$ and $d$ quarks are fixed
rather well by the infrared tail of the light $D$-function and the value of
$R_\tau^V$. If we take for the parameter
$M_0^{u,d}$ in the function $m=m(p^2)$ the best fit value $260$~MeV and
vary $M_0^s=400$--$500$~MeV, we get
\begin{equation}\label{a_mu-our}
a_\mu^{\rm{had}}=(698\pm13)\times10^{-10}.
\end{equation}

Alternative ``theoretical''  values of $a_\mu^{\rm{had}}$ are extracted from
$e^+e^-$ annihilation and $\tau$ decay
data: $(696.3\pm6.2_{{\rm exp}}\pm3.6_{{\rm rad}})\times10^{-10}$
($e^+e^-$-based) \cite{Davier:03}, which is
1.9$\sigma$ below the BNL experiment \cite{bnl}; $(711.0\pm5.0_{{\rm exp}}
\pm0.8_{{\rm
rad}}\pm2.8_{SU(2)})\times10^{-10}$ ($\tau$-based) \cite{Davier:03}, which
is within 0.7$\sigma$  of experiment; and
$(693.4\pm5.3_{{\rm exp}}\pm3.5_{{\rm rad}})\times10^{-10}$ ($e^+e^-$-based)
\cite{Hocker:04}, 2.7$\sigma$ below
experiment. An even lower value $(692.4\pm5.9_{\rm exp}\pm2.4_{\rm rad})
\times 10^{-10}$ is given by \cite{Hagiwara:04}.
The quantity $a_\mu^{\rm{had}}$ is rather sensitive to the light
quark mass parameters,  which are known only with large
uncertainties. For this  reason our estimations at this stage cannot
give a preference to one or another of the above-mentioned
fits to experimental data.

\subsection{Hadronic contributions to $\Delta\alpha$}
Consider the hadronic correction to the electromagnetic fine structure constant
$\alpha$ at the $Z$-boson scale. The
evolution of the running electromagnetic coupling is described by
\begin{equation}\label{alpha-s-evolution}
\alpha(s)=\frac{\alpha(0)}{1-\Delta\alpha_{\rm{lept}}(s)
-\Delta\alpha_{\rm{had}}^{(5)}(s)-\Delta\alpha_{\rm{had}}^{\rm{top}}(s)}.
\end{equation}
The leptonic part $\Delta\alpha_{\rm{lept}}(s)$ is known to the
three loop level,
$\Delta\alpha_{\rm{lept}}(M_Z^2)=0.03149769$ \cite{Steinhauser:98}.
It is conventional to separate the contribution
$\Delta\alpha_{\rm{had}}^{(5)}(s)$ coming from the first five quark flavors.
The contribution of the $t$-quark is
estimated as $\Delta\alpha_{\rm{had}}^{\rm{top}}(M_Z^2)=-0.000070(05)$
\cite{Kuhn-Steinhauser:98}.

The quantity $\Delta\alpha_{\rm{had}}^{(5)}(s)$ at the $Z$-boson scale can be
represented in the form of the
dispersion integral (\ref{Delta-alpha-R}). The total function $R(s)$ is
\begin{equation}\label{R(s)-sum-R_f}
R(s)=3\,\sum_fQ_f^2{\cal{R}}_f(s),
\end{equation}
where $Q_f$ is the quark electric charge of flavour $f$.
For the calculation of $R(s)$ we use (\ref{R(s)-sum-R_f}) with five quark
flavors $f=u,d,s,c,b$. Varying the parameters
as has been described above and using $m_c=1.3$--$1.5$~GeV, we get
\begin{equation}\label{Delta-alpha-result}
\Delta\alpha_{\rm{had}}^{(5)}(M_Z^2)=(278.2\pm3.5)\times10^{-4}.
\end{equation}
This value is to be compared with predictions extracted from a wide
range of data describing $e^+e^-\to$ hadrons
\cite{Hagiwara:04}:
\begin{equation}\label{Delta-alpha-exper}
\Delta\alpha_{\rm{had}}^{(5)}(M_Z^2)=(275.5\pm1.9_{\rm{expt}}
\pm1.3_{\rm{rad}})\times10^{-4}.
\end{equation}
We see that our result (\ref{Delta-alpha-result}) is consistent with
previous theoretical/experimental evaluations, with comparable uncertainties.

The relative error in (\ref{Delta-alpha-result}) is substantially less
than the error that appears in the quantity
$a_\mu^{{\rm{had}}}$ and therefore one can obtain a more exact result.
In comparison with the $a_\mu^{{\rm{had}}}$ result, where
the contribution of the $c$-quark was about $2\,\%$, now it is about $30\,\%$.
The contribution of the $b$-quark is about
$5\,\%$ and the relative contribution of the $t$-quark is a fraction
of a percent.

\section{Conclusions}
A method of performing QCD calculations in the nonperturbative domain has been
developed. This method is based on the
analytic approach to QCD, in which there are no unphysical singularities, and
takes into account the  summation of
threshold singularities and the involvement of
nonperturbative light quark masses.

The following quantities have been analysed:
the inclusive $\tau$-decay characteristic in the vector channel, $R_\tau^V$;
the light-quark Adler function, $D(Q^2)$;
the smeared $R_\Delta$-function;
the hadronic contribution to the anomalous magnetic moment of the muon,
$a_\mu^{{\rm{had}}}$;  and
the hadronic contribution to the fine structure constant,
$\Delta\alpha_{\rm{had}}^{(5)}(M_Z^2)$.
We have demonstrated that the proposed method allows us to describe these
quantities rather well.

\acknowledgments

 It is a pleasure to thank Prof.~D.V.~Shirkov for interest in the work,
support and useful
discussion. We express our gratitude to Drs.~A.E. Dorokhov, S.B. Gerasimov,
A.V. Efremov, and O.V. Teryaev,  for valuable discussions and helpful remarks.
This work was supported in part by the International Program of Cooperation
between the Republic of Belarus and
JINR, the Belarus State Program of Basic Research ``Physics of Interactions,"
and RFBR grant No.~05-01-00992, and NSh-2339.2003.2.
K.A.M. is grateful to the Physics Department of Washington University for
its hospitality and support, and to the US Department of Energy for grant
support.

\end{document}